\documentclass[12pt]{iopart} 
\usepackage{amsthm}
\usepackage{amssymb}
\usepackage{graphics,graphicx}
\usepackage{epsfig}
\usepackage{multicol}
\usepackage{color}
\usepackage{bm}
\renewcommand{\theequation}{\thesection.\arabic{equation}}

\newcommand{\mpl}{m_{pl}}
\newcommand{\fnl}{f_{\rm NL}}
\newcommand{\N}{\mathcal{N}}

\begin{document}
\title{Towards an Observational Appraisal of String Cosmology}
\author{David J. Mulryne$^1$ and John Ward$^2$}
\address{$^1$ Astronomy Unit, School of Mathematical Sciences, Queen Mary University of London, Mile End Road, London, E1 4NS, UK}
\address{$^2$ Department of Physics and Astronomy, University of Victoria, Victoria, BC, V8P 1A1, Canada}
\ead{d.mulryne@qmul.ac.uk; jwa@uvic.ca}

\begin{abstract}
We review the current observational status of string cosmology when confronted with experimental datasets. We begin by defining common observational parameters and discuss how they are determined for a given model. Then we review the observable footprints of several string theoretic models, discussing the significance of various potential 
signals. Throughout we comment on present and future prospects of finding evidence for string theory in 
cosmology, and on significant issues for the future.
\end{abstract}

\section{Introduction}
Thanks to important advances in experimental astrophysics,
the past two decades has seen modern cosmology 
become a high precision, data rich science (see for example 
\cite{Astier:2005qq,Perlmutter:1998np, Riess:2004nr, Riess:1998cb,Komatsu:2010fb,Spergel:2006hy, Spergel:2003cb, Tegmark:2003ud, Tegmark:2006az}).
Observations suggest that our universe is consistent with a flat, so called `$\Lambda$CDM 
universe', and have confirmed that
our favoured theory for the origin of structure, cosmological
inflation \cite{inflation} is well supported by the data.

Given the extremely high energy scales present in the early universe, and
huge distances probed by large scale
cosmological evolution, cosmological observables may be sensitive to Planck scale physics.
String theory is a leading contender for a theory of the ultraviolet
(UV) completion of quantum field theory and gravity, and 
hence one within
which Planck scale effects can be addressed. It is
natural, therefore, to ask what the consequences
of cosmology are for string theory,
and visa versa. Thus the notion of `string cosmology' \cite{Lidsey:1999mc} was born.

Ideally string cosmology is the \emph{direct} application of string theory to understanding 
the evolution of the universe. 
Then, through comparison with data, we 
might hope to experimentally test the theory.
In practice, however, this is too ambitious and generally models 
are constructed using ideas and intuition from string theory, 
which are then confronted with observation. Although such a program
has been remarkably successful, the inherent stringiness of the models 
is seldom explicitly present, and it is 
difficult to argue that constraining models is a 
direct probe of string theory itself.

On the other hand, particular theories can be sensitive to UV physics, even if not probing it
directly, and certain self-interaction potentials for the inflaton can
arise in string motivated models which are rather unlikely to arise from pure field
theory constructions. Moreover, the self consistency of string inflation models 
can certainly be probed by observation. 
Furthermore, while a less developed area of research,
string cosmology extends beyond the inflationary epoch.
It may, for example, offer convincing alternatives to inflation, which flow more directly from the UV complete nature of the
theory. It is also possible to generate cosmic {\it super}strings which 
could be directly detected by observation. String theory may even have a 
role to play in understanding why our universe is accelerating today.

With all this in mind, the purpose of the present review is to ask how far we
have come in our quest to probe string theory using cosmology, and to
address questions such as: will we ever see observational 
signatures of string theory in
cosmology? And what are the most promising signals to look for? 

We structure the review as follows. In Section \ref{sec:Obs}, we review how 
inflationary models are confronted with observation, and the strength of 
present and future constraints, as well as discussing other observations relevant to 
probing string theory. In Section \ref{sec:Test}, we discuss how 
string theory models are being tested by observation, 
discussing what would constitute 
evidence of string theory in light of the 
issues of naturalness and robustness of models, and review a number of inflationary models 
and others together with 
their observational predictions. Finally we conclude in Section
 \ref{sec:Discuss} by highlighting what the key issues are for the future.

\section{Observations and discriminators of early universe models}\label{sec:Obs}

There exists an extremely 
well developed framework for 
determining how well a given inflationary model (or alternative)  
fits experimental data, which we will 
now review.  We then discuss current 
and future observational 
constraints, as well as other observations of interest for string cosmology. 

\subsection{Discriminators of the very early universe}
\label{subsec:dis}
Inflation is the accelerated expansion of spacetime; during which 
quantum fluctuations of the metric and matter are `stretched' to large scales, and subsequently 
become the origin of cosmic structure (for useful reviews see for example 
\cite{Lyth:2009zz, Lyth:1998xn, Guth:2005zr, Lidsey:1995np}). 
The fundamental scalar quantity is the primordial curvature perturbation 
$\zeta$ and tensor fluctuations are parametrised by their amplitude, $T$ 
(see for example \cite{Bardeen:1983qw,perts}).
Isocurvature perturbations may also be produced and persist after inflation, but are not inevitable 
and are disfavoured by current data.  
In the absence of isocurvature perturbations,  
a given wavenumber, $k$, of $\zeta$  
is conserved once stretched to super-horizon scales, $k<aH$, where $a$ 
is the scalar factor and $H=\dot{a}/a$ \cite{Bardeen:1983qw,Lyth:1984gv,Wands:2000dp,Rigopoulos:2003ak,Lyth:2004gb}. 
The stochastic properties of these perturbations are 
probed by observation. The full power spectrum, parametrising the two-point 
function, can be a powerful tool, but is generally 
parametrised about some pivot scale, $k_*$ in terms of a number of key parameters.
The first are the square of the amplitude of the 
scalar and tensor modes, denoted ${\cal P}_\zeta^*$ and 
${\cal P}^*_T$ respectively, 
which lead to the ratio $r={\cal P}^*_T/{\cal P}^*_\zeta$. This parameter 
is important because a 
detection would directly probe
the energy scale of inflation.
The next key parameters 
are the spectral tilts; 
$n_s(k)-1 = d \ln {\cal P}_\zeta / d \ln k|_*$ for scalar 
perturbations, and $n_T(k) = d \ln {\cal P}_T/ d \ln k|_*$ for 
tensors.  One can then continue to define a running, the derivative of the tilt with respect to $\ln k$, 
and higher derivative parameters if required. 

Further information is available 
from studying statistics beyond the two-point 
function. The three-point function, which vanishes 
for Gaussian perturbations, is parametrised by the bispectrum, $B_\zeta (k,k',k'')$, 
generally normalised by the square of the power spectrum to give the 
reduced bispectrum or $\fnl(k,k',k'')$ parameter. One can then continue to the trispectrum 
(four-point function), running of $\fnl$
and so on. Currently meaningful constraints 
exist only for the subset of parameters $\{r, n_s, \fnl\}$ and the running of $n_s$.

Single field inflationary models are the most widely studied, because of their 
simplicity, and are characterised by the generalised action
\begin{equation}
\label{action}
S = \int d^4 x \sqrt{-g} \left[\frac{\mpl^2}{2} R + P(\phi,X) \right ]\,,
\end{equation}
where $R$ is the Ricci scalar, minimal coupling has been assumed and $X = -\frac{1}{2}g^{\mu \nu} \nabla_\mu \phi \nabla_\nu \phi$.
For inflation driven by this action it has been found that \cite{Garriga:1999vw}
\begin{equation}
{\cal P}_\zeta = \frac{1}{8 \pi^2 \mpl^2} \frac{H^2}{c_s \epsilon}\,,\hspace{0.4cm} {\cal P}_T = \frac{2}{\pi^2} \frac{H^2}{\mpl^2}  
\end{equation}
where $\epsilon = -\dot{H}/{H^2}$, $c_s$ is sound speed of scalar fluctuations, $c_s^2 = P_{,X}/(P_{,X}+2XP_{,XX})$, 
and all expressions are evaluated at the scale $k^*=aH/c_s$. 
We note that accelerated expansion 
requires $\epsilon <1$, and $c_s =1$ corresponds to an canonical scalar field with 
modes travelling at the speed of light, for which
$\epsilon \approx \frac{m_{pl}^2}{2} (V'/V)^2$. 
Moreover one finds
\begin{equation} 
r=16 c_s \epsilon \,,~~ (1-n_s)=2\epsilon +\frac{\dot{\epsilon}}{\epsilon H} + \frac{\dot{c_s}}{c_s H}\,,~~ n_t=-2\epsilon
\end{equation}
and for equilateral triangles $k \sim k' \sim k''$ \cite{Maldacena:2002vr,Seery:2005gb,Chen:2006nt}
\begin{equation}
\label{fnl}
\fnl^{\rm eq}=\frac{5}{81} \left ( \frac{1}{c_s}-1-2\Lambda\right) - \frac{35}{108} \left (\frac{1}{c_s}-1\right)\,,
\end{equation}
where, $\Lambda = (X^2 P_{,XX}+ \frac{2}{3}X^3 P_{,XXX})/(X P_{,X}+2X^2 P_{,XX})$. All shapes of 
$\fnl$ are negligible for canonical single field models with $P=X$. 

To predict the observational parameters from a given model of inflation, we must find the 
values that parameters ($\epsilon$, $c_s$ etc.) took when $k^*=aH/c_s$. 
This is dependent on the number of e-folds ${\cal N^*} = \ln(a_{\rm end}/a^*)$, which in 
turn depends on post inflationary physics. 
A reasonable range is ${\cal N^*} \approx 54 \pm 7$ \cite{Liddle:2003as}, 
but values \emph{well} outside this range 
are possible. Since the observational footprint of any given model will be sensitive to $\cal N^*$, 
to properly compare a given model with observation one must generally determine 
its \emph{post-inflationary}
behaviour. Unfortunately this is not known in most models.

Space will not permit a careful review of how observational 
predictions are made for every string cosmology 
model we will discuss, so here, 
following \cite{Alabidi:2008ej,Alabidi:2010sf} (where the reader can turn for a fuller discussion of observational constraints), 
we include two illustrative examples of canonical models 
(which cover a large number in the literature):
\begin{equation}
\label{examples}
{\rm(a)}~V= V_0 \left [1-\left ( \frac{\phi}{\lambda} \right)^p\right]\,,~~~~ {\rm (b)}~V= V_0 \left (\frac{\phi}{\lambda}\right)^{p}\,,
\end{equation}
where $V_0$, $\lambda$ and $p$ are constants. Assuming the potential 
maintains this form over the entire inflationary evolution, 
and using the approximate expression $d {\phi}/ d {\cal N} \approx -\sqrt{2 \epsilon \mpl^2}$, 
which follows when $\epsilon\ll1$, one can find the field value and thus all relevant 
quantities ${\cal N^*}$  e-folds before the 
end of inflation (defined as $\epsilon = 1$). 
Potential (a) represents a small field model for which the range of field 
values traversed during inflation is $|\Delta \phi| < \mpl$, and (b) represents 
a large field 
model where $|\Delta \phi| > \mpl$, raising the usual issue 
of corrections to the potential. Following the procedure outlined, for (a) one finds that $r$ is negligible (typical 
of small scale models since one can in general express $r=8 (d\phi/d{\cal N})^2/\mpl^2$ \cite{Lyth:1996im}), 
and $n_s = 1-2 \left( \frac{p-1}{p-2}\right)\frac{1}{\cal N^*}$ (the case of $p=-\infty$ 
corresponds to the potential $V=V_0 (1- e^{-a \phi/\mpl})$, and $p=0$ to $V=V_0(1+A\ln(\phi/B))$). 
For (b) one finds $1-n_s=\frac{2+p}{2 {\cal N^*}}$ and 
$r=\frac{8[{\cal N^*}(1-n_s)-1]}{\cal N^*}$ (where we have considered $p>0$).  
The relation between $n_s$ and $r$ is important because the corresponding parameter 
space is well constrained.  Note, however, that such simple expressions follow from the simplicity  
of the potential. Were there additional (potentially unknown) 
terms, such simple relations would not exist. There are, therefore, two
key lessons of this discussion. Firstly observables depend on ${\cal N}^*$, which we don't 
know apriori and is not itself an observable. And secondly, simple relations between 
parameters are possible but will be spoilt if the form of the potential is altered by further terms
arising from quantum corrections, and which may introduce new parameters.

Further complications arise if more than one 
field is light at horizon crossing, since isocurvature modes will be produced. No 
observational evidence that such modes existed has been found, but were it to be it 
would rule out single field inflation. 
When isocurvature modes are present, the curvature perturbation and 
its statistics may evolve on super-horizon scales during inflation 
if the field space path curves \cite{Gordon:2000hv}.
Even if isocurvature modes decay before or during reheating, 
a curved path during inflation will alter the relation between observational 
parameters and the value of $\epsilon^*$ etc. at horizon 
crossing (even if the path only curves after modes around the pivot scale exit the horizon). 
The best developed method to account for this is the $\delta {\cal N}$ formalism 
\cite{Starobinsky:1986fxa,Sasaki:1995aw,Lyth:2005fi}. Space restricts 
us from providing the full details, but for canonical inflation the amplitude of the 
power spectrum is given 
by ${\cal P}_\zeta = {\cal N}_{,i}{\cal N}_{,i} H^2/(4 \pi^2)$, 
the tilt by $1- n_s = 2\epsilon^* - 2\dot{\phi_i}^* {\cal N}_{,i j} {\cal N}_{,j}/(H^* {\cal N}_{,k} {\cal N}_{,k})$, 
and the most important contribution to the reduced bispectrum 
in the squeezed limit ($k\sim k'<<k''$) is
$\fnl^{\rm loc} = 5/6 {\cal N}_{,i} {\cal N}_{,i j} {\cal N}_{,j}/({\cal N}_{,k} {\cal N}_{,k})^2$, 
which can be large for certain models. 
Here ${\cal N}$ is the number of e-folds from 
horizon crossing to a constant energy density hypersurface once the evolution has become adiabatic, 
roman numerals label the ${\cal M}$ light fields, and the 
subscripts denote derivatives with respect to changes in the field values 
at horizon crossing. In multi-field models, therefore, simple relations for quantities 
such as $n_s$ are only available for the simplest trajectories, and moreover, a ${\cal M}-1$ 
dimensional surface now leads to any given ${\cal N}^*$. 

\subsection{Observational constraints and other signatures}

Observations of the CMB constitutes the most important tool at our disposal to constrain 
the defined parameters. Normalisation of the CMB anisotropy requires ${\cal P}_\zeta =2.42 \times 10^{-9}$ \cite{Komatsu:2010fb}.  
Precise constraints on the observational parameters depend on how many parameters are included in the 
statistical analysis, and what other data sets are included.  For example, at the $68\%$ confidence level, if the running of the scalar spectral 
index and $r$ are 
assumed to be zero, the WMAP-$7$ \cite{Komatsu:2010fb} data alone implies 
$n_s \approx 0.967 \pm 0.014$. If $r$ is also included the 
data gives $n_s \approx 0.982\pm 0.02$ and $r<0.36$ (at the $95 \%$ confidence level), while allowing for running of the scalar spectral 
index leads to $n_s \approx 1.027 \pm 0.05$ and  $d n_s / d \ln k \approx -0.034\pm 0.026$ ($r$ taken to be zero). 
Moreover, it is important to recognise 
that the analysis assumes the absence of other contributions to density fluctuations, such as cosmic (super)strings. 
When these were included, a recent study found a blue spectrum ($n_s>1$)  
could be accommodated \cite{Bevis:2007qz} ($n_s=1.00\pm0.03$ with a maximum $11\%$ 
contribution to power from cosmic strings). 
An important outcome, therefore, is to recognise that while statements 
such as the WMAP data favours a red ($n_s<1$) spectrum are common, this is highly analysis dependent. 

WMAP also
constrains non-Gaussianity, with $-10<\fnl^{\rm loc}<74$ 
and $-214 < \fnl^{\rm eq} <266$ at the $95\%$ confidence level. 

The Planck satellite \cite{:2011ah} currently taking data will hopefully improve these constraints considerably, 
and in particular, in the absence of a detection one expects limits of roughly $|\fnl|<5$, and $r<0.05$ and error bars 
on $n_s$ at least an order of magnitude better than WMAP.  
The proposed CMBPOL mission \cite{Baumann:2008aq} (designed specifically to 
look at the polarisation of the CMB) could probe down to $r<0.01$. 
There are other exciting 
future possibilities, such as observation of 21cm radiation \cite{Khatri:2009aw}, which probes the structure of the universe 
during the cosmic dark ages before re-ionisation, and could give limits of $|\fnl|<1$. Other important observations which 
constrain primordial perturbations over different scales are the various surveys of 
large scale structure, which are 
often combined with WMAP data, and in particular can give constraints on the running of parameters (see for example \cite{Dalal:2007cu}).

CMB polarization is also an important discriminator in its own right. Scalar perturbations 
generate only E-modes, whilst tensor perturbations
generate both E and B-modes. Vector perturbations also generate B-modes (the E-mode being negligible with respect to the B-mode),  
and while highly 
 suppressed during inflation are sourced by cosmic strings. Thus the detection of B-modes would  
automatically lead to exciting new knowledge about the universe.  
One caveat is that we must assume there is no axionic
coupling to the photon through terms of the form $ \sigma a F \wedge F$, since this can rotate the 
E-mode into a B-mode with mixing angle given by 
$\Delta \theta \sim \sigma \Delta a $. Current WMAP bounds on this angle at the $68 \%$ confidence level are
\begin{equation}
\Delta \theta = -1.1 \rm{deg} \pm 1.4 \rm{deg} (\rm{stat}) \pm 1.5 \rm{deg} (\rm{syst})
\end{equation} 
which is consistent with it vanishing, but further observations 
are clearly required \footnote{Such a coupling can arise in string models through the inclusion of Wess-Zumino (WZ) terms.}.

Observational cosmology is an incredibly rich field, and important data for string cosmology may 
lurk in numerous other observations. As we will discuss, gravitational lensing - both strongly lensed images
and weak lensing - could contain information about cosmic strings, as could micro lensing surveys \cite{Das:2011ak}. 
Moreover, data is available on the peculiar velocities of clusters of galaxies, which probe the laws of gravity on the 
largest scales \cite{Feldman:2009es}.  Not to mention the improving data from supernovae which played the crucial role in determining the need for 
a dark energy component (acting like a cosmological constant) 
to accelerate the universe today \cite{Guy:2010bc}. The potential evolution of dark energy will be 
a key focus of future observations. Finally, we note there is potential for 
the direct observation of gravitational waves by ground or space based interferometers (LIGO and LISA) \cite{:2010yba,Babak:2010ej}.  Though we will 
likely have to wait many years before an experiment has the sensitivity to be relevant to inflationary gravitational 
waves (BBO) \cite{BBO}, constraints relevant to cosmic strings already exist \cite{Wyman:2005tu}.

\section{Testing string theory using cosmology}\label{sec:Test}

The overall goal of the string cosmology program is to use cosmology as a 
testing ground for string theory. The twin aims are to understand 
whether and how string theory constructions can explain observed  
properties of the universe, and, more excitingly, to 
determine whether there might be a \emph{signal} of string theory in observations. 
The holy grail would be an observational confirmation through the
direct detection of something genuinely `stringy'.
It is possible, perhaps even likely, though 
that there will be no smoking gun, rather, 
evidence for string theory might come in a less dramatic form; for example 
by providing a natural, convincing explanation for something
already observed but not properly understood. An example would be a truly 
compelling model of inflation 
(or an alternative).
It may transpire, however, 
that all we can achieve
is consistency of models with observables, and nothing more.
Yet even if this were so, a well motivated string inflationary model 
which passed successive observational tests of this type, at the
expensive of other models failing, could come to be seen as evidence of
string theory itself. Likely such evidence would need to be augmented by
other observations. For example if such a model additionally predicted
cosmic superstrings and some evidence of cosmic strings was found, 
it would become doubly appealing.

In light of these points, as we discuss aspects of string cosmology -- and models of the early universe in particular -- 
we will keep three issues in mind.
The first is naturalness of the model. This is a hard concept to make precise but, for 
example, a model for which a vast range of parameters or initial conditions is allowed, but for which 
only a vanishingly small range leads to a consistent cosmology, could be viewed as unnatural.  
The second is testability. Is the model sufficiently well developed that 
all its parameters are derivable and constrained by observation? Or does it require   
elements that are plausible, but not yet consistently implemented?  Ideally to 
test models we should have more observable parameters than model parameters, if not 
we can only probe combinations of model parameters. 
Finally a model could be predictive, in the limited sense that it leads to a `stringy' prediction, which 
could not come from, or would be hard to produce, in the absence of stringy physics.

\subsection{String Inflation}
First we consider string inflationary models. We aim to review a representative selection, discuss 
how they confront observation and attempt to address how well they measure up against 
the issues of naturalness, testability and predictivity.
We will be particularly interested in those that include concrete calculations of 
world-sheet or loop corrections, since these are both a sign 
of the maturity of a model, and will likely lead to reliable signals and consistency 
relations. We will also focus on observable features which seem 
to appear more naturally in a string theory setting.  

Before embarking on this path, we first note that while many parameters appearing 
in a model, such as background fluxes, are inherently stringy, we cannot   
measure them directly cosmologically because we restrict ourselves to an effective inflaton action.
Such parameters are absorbed by field redefinitions, giving rise 
to additional degeneracies, and it is difficult to argue that by probing a given 
model we are truly probing the original stringy parameters. Moreover, we adopt a critical position that 
although supergravities are the low energy limits of string
theory, an observable signature within supergravity is not (in itself) proof of a 
string theory signal. 

{\bf \it Modular inflation.}
Most work in this area has focused on flux compactification of type IIB supergravity. Such compactifications 
preserve at least one, but in general many, massless moduli due to the no-scale structure of the classical theory.
These fields can acquire a mass through non-perturbative contributions to the superpotential, such as those arising
from gaugino condensation on wrapped $D$-branes. However such terms can only be calculated explicitly in string theory
in a handful of cases, since they typically depend on moduli that have been integrated out of the theory.
The first attempt to fix all moduli was the KKLT construction \cite{Kachru:2003aw} which considered 
one complex modulus. The resulting potential requires an additional uplifting term provided by an
anti-$D3$ ($\bar{D3})$ brane at the tip of the warped throat, to obtain a $dS$ vacuum, and doesn't 
lead to a consistent inflationary scenario, but the basic procedure underlies many other models, 
and including more than one moduli can lead to viable inflationary models.

One such scenario, tree level in loops but including world-sheet corrections,
is Racetrack inflation on the $\mathbb{CP}^4[1,1,1,6,9]$ Calabi-Yau (CY) three-fold \cite{BlancoPillado:2006he,BlancoPillado:2009nw}. 
The complicated `Racetrack' potential arises from competition between competing terms in the non-perturbative
superpotential, with two scalar fields driving inflation \cite{BlancoPillado:2004ns}. Inflation is possible with 
particular initial conditions, but not generic,  
on this potential
and satisfying the WMAP normalisation of the power spectrum proves challenging.
The authors restricted themselves to variations of the constant term in the superpotential 
$(W_0)$, and found that maximising the manifold volume led to $n_s \sim 0.95$ for a straight trajectory evolving from 
the saddle point and emulating a small field model, with negligible $\fnl$ and $r$. 
While a useful proof of principal, meaningfully probing the parameter space 
using observations for this model would be extremely difficult.

Another model of interest is based on the large volume scenario (LVS) 
\cite{Conlon:2005ki, Conlon:2005jm} with one or more of the geometric 
moduli identified as the inflaton \footnote{This is a K\"ahler modulus in IIB and a complex structure modulus in IIA.}. 
The world-sheet corrections ensure that the volume is stabilised at exponentially large values after inflation.
There are several different models in this class including K\"ahler inflation, Roulette inflation \cite{Bond:2006nc} and Fibre inflation \cite{Cicoli:2008gp} 
which all have exponential
potentials. Both K\"ahler and Roulette inflation lead to small tensor models $r \sim \mathcal{O}(10^{-10})$, with the simplest K\"ahler model having 
the form $V=V_0 ( 1 - e^{-a\phi})$, and hence leads to  
$n_s = 1-2/\N^*$ (this corresponds to the $p=-\infty$ small field example of Section \ref{subsec:dis}). In Roulette models
the inflaton is associated with a classical trajectory through field space (perpendicular
to the isocurvature trajectory) \cite{Bond:2006nc}.
 However inflationary trajectories typically have $\epsilon^* \sim 0$, 
which suggests conflict with the WMAP data.
In the more general multi-field scenario, results indicate a larger red-tilted power spectrum with $|\fnl| < 0.1$.
Fibre inflation consists of a $K3$-manifold fibred over a $\mathbb{CP}1$, allowing for the explicit inclusion of 
one-loop corrections. These corrections are quantified by $R = 16 AC/B^2 $, where $A, C$ are terms explicitly arising from loop effects. Sufficient 
e-folds are obtained for small $R$, with
 $\N =60$ occurring for $R = 3 \times 10^{-5}$. Therefore in the limit that $R \to 0$ one finds
the model independent footprint\footnote{Meaning that the observables only depend on the slow roll parameters.} $r \simeq 6 (n_s-1)^2$ which is 
 within current WMAP bounds. In the opposing regime we find $r \simeq (32/3) R^{2/3}, n_s \simeq 1 - 4 R^{2/3}$
which implies $r \leq 0.01,  n_s \leq 0.996$ at $\N =60$. A two-field model was also constructed in this class with similar results \cite{Cicoli:2008gp}.

These large volume models 
are well motivated, considered natural, and are testable at least for 
single field constructions, though they do not yet predict any genuinely stringy signatures. 
One should note that loop corrections are not universal, although their general
form is known \cite{Berg:2007wt}, and must be computed for each CY.
The prototypical case is $\mathbb{CP}^4[1,1,1,6,9]$, where the leading order perturbative and non-perturbative world-sheet corrections, and
the one and two-loop terms are known \cite{Misra:2007cq}. 
The non-perturbative corrections ensure a $dS$ vacuum without the need for $\bar{D3}$ branes, however
the loop corrections were argued to be \emph{sub-dominant} with respect to the world-sheet corrections.
The inflaton here \cite{Misra:2007cq} is a linear combination
of NS-NS axions, and inflation occurs at a saddle-point where $\cal N$ depends explicitly on the degree of the genus-zero
holomorphic curve. Subsequent work computed cosmological observables, which are sensitive to the volume $(\mathcal{V})$ and the $D3$-instanton number $(n^s)$. 
Favourable scenarios require $\mathcal{V} \gg 1, n^s \sim 10$, which yields $|f_{nl}| \sim 10^{-2}, r \sim 10^{-4}$ and
$|n_s-1| \sim 10^{-3}$ \cite{Misra:2007cq}. 
Although one can compute various (soft) susy-breaking masses and Yukawa couplings, which are themselves expected to be experimentally constrained,
direct cosmological observation will require the inclusion of loop corrections to distinguish these models from field theory. 

\subsection{Brane inflation models} 
$D$-branes play a key role in modern string theory, so it is natural to consider their cosmological consequences. These branes 
are described by non-linear Dirac-Born-Infeld (DBI) theory.
The most popular class of cosmological models exist in IIB, where a mobile $D$ brane travels relativistically down 
a warped throat towards attractive $\bar{D}3$-brane charge \cite{Alishahiha:2004eh}. In the simplest case 
$P(\phi, X)$ is of the form
\begin{equation}
P=-T(\phi) \sqrt{1-2X T(\phi)^{-1}} + T(\phi)  - V(\phi)
\end{equation}
with $T$ the warped brane tension and $V(\phi)$ the scalar potential.
The non-linear nature of DBI-inflation ensures that $c_s$ can become small, leading to large equilateral $\fnl$ (Eq.\ref{fnl}). 
These models are potentially testable \cite{Lidsey:2007gq}, predictive and moreover the speed limit imposed by the warped 
geometry was originally thought to make 
inflation extremely natural. Combined, these features have 
lead to significant interest in such models. 
The brane, however, can only travel a finite distance $\Delta \phi$ due to the finite length 
of the warped throat \cite{Chen:2006hs,Baumann:2006cd}, 
which translates to an upper limit on $r$, as discussed in Section \ref{sec:Obs}. 
When implemented with the above $D3$-brane action, assuming 
the simplest throat constructions, 
one finds \cite{Baumann:2006cd,Lidsey:2006ia}
\begin{equation}
r_{*} < \frac{32}{\N \N_{eff}^2}(c_s P_{,X})_{*}
\end{equation}
where $30<\N_{eff}<60$ is expected. 
Moreover, when combined with Eq.(\ref{fnl}) and other observational constraints, 
this implies $\fnl$ would have to be outside the 
WMAP bounds \cite{Lidsey:2006ia}. A result which is independent of
the scalar potential.
While this is disappointing, it highlights that we are now genuinely 
able to confront string theoretic models in an increasingly powerful way. Before 
calculations of $\fnl$ were performed, 
and the WMAP constraints available, this model would have appeared viable at the level of the power spectrum. Now 
despite its appeal, in its simplest form at least, it can be ruled out. 
Note that small field DBI-inflation may evade this bound 
while still leading to 
an $\fnl$ signal within reach of Planck (see for example \cite{Bean:2007eh}).
Moreover complex models can be constructed which 
relax the above bounds and include, angular modes, wrapped branes, 
multiple branes and multi-field theories 
\cite{Becker:2007ui} - though the 
issue of naturalness must be raised in this context. 
One such proposal, independent of the scalar potential using 
multiple/wrapped branes, links the tensor-scalar
ratio directly to $\fnl$ via
\begin{equation}
r_{*} < - \frac{5}{\mathcal{N}^2_{eff}} \frac{f_{nl}}{(n-1)\sqrt{N}}
\end{equation}
where $N$ is the $D3$-brane charge of the $AdS_5 \times X_5$ geometry, and 
$n$ is the number of branes. Such a model is clearly ruled out
if $f_{nl}$ is observed to be zero, or has positive sign. In the case of wrapped  
$D(3+2k)$-branes, it was found that the
backreaction becomes more important for higher $k$. The wrapped $D7$-brane bound becomes
\begin{equation}
\frac{(1-n_s)}{8} < r < \frac{2 16 \pi}{3^4} \frac{K^4 P_s^2}{\rm{Vol}(X_5) (\Delta \N)^6}\left(1 + \frac{1}{3 f_{nl}} \right)
\end{equation}
which can be satisfied for $K \ge 46$, where $K \in \mathbb{Z}$ is the NS-NS flux at the tip of the throat.
There are many possible extensions, and it is likely to be a major area of inflationary model building for
some time to come.
 
Another open string model embedded into IIB compactifications is that of $D3/D7$-brane inflation \cite{Dasgupta:2002ew, Haack:2008yb}, 
where the compact manifold is $K3 \times T^2/Z_2$.
The inflaton is related to the distance between the two
types of brane on the orbifold, and its potential is generated by the presence of non self-dual 
flux on the $D7$-brane. This generates a non-zero D-term
potential with Fayet-Illioupolous (FI) parameter $\xi$. The resulting mechanism is a stringy
version of hybrid inflation, and inclusion of loop effects leads to the footprint \cite{Haack:2008yb}
\begin{equation}
n_s  = 1 -\alpha \left(1 + \frac{1}{(1 - e^{-\alpha \N})} \right), \hspace{1cm} \alpha = \frac{4 m^2}{g_s^2 \xi^2}
\end{equation}
An interesting consequence of many brane models is the creation of cosmic superstrings, which is perhaps the most 
predictive of all possibly observable stringy physics. 
This model is a particularly interesting example for which strings with a tension spectrum $G \mu \sim \xi/4$ are produced, and 
which could be be used to constrain or fix the FI parameter. 
Under the assumption that cosmic strings contribute at most $\mathcal{O}(11\%)$ to the power spectrum, this implies $r < 10^{-4} g_s^2$, which is vanishingly
small for perturbative strings. Increasing the scalar mass tends to suppress the cosmic string contribution, but shifts $n_s$ further towards unity, and 
towards unfavoured WMAP values. This is another interesting example of how combinations of observables can probe or constrain a model.

One interesting recent development in brane inflation models, 
has been to consider corrections to the inflaton potential, for  
brane models in which the brane is moving non-relativistically,  
from compactification effects in the throat \cite{Kachru:2003sx, Baumann:2010sx}. Such a calculation has been
done for a class of the simplest $D3$/$\bar{D}3$ models discussed above.
At leading order the inflaton potential is generated by the Coulombic interaction
between these branes, however the corrections due to a single angular mode can be included resulting in the following potential
\begin{equation}
V(\phi) = V_0 (\phi) + M_p^2 H_0^2 \left(\left( \frac{\phi}{M_p}\right)^2 - a_{\Delta} \left( \frac{\phi}{M_p}\right)^{\Delta}\right), \hspace{0.6cm} 
a_{\Delta} =  c_{\Delta} \left( \frac{M_p}{\phi_{UV}}\right)^{\Delta}
\end{equation}
where $V_0 (\phi)$ generally includes all terms that yield negligible corrections to $\eta$.
Note that $\Delta$ corresponds to the eigenvalues of the compact Laplacian, and the smallest eigenvalue ($\Delta  =3/2$) is expected to dominate. However
if symmetries forbid the existence of such a term, the next possible contribution comes from quadratic modes $\Delta = 2$. Such a model is relatively 
generic in that it relies on the computation of the Laplacian in the non-compact throat, rather than on the full details of the compact space.
The potential in the case of $\Delta = 2$ has been well studied since it is of the form $V(\phi) = V_0(\phi) + \beta H_0^2 \phi^2$. Slow
roll inflation requires $\beta \ll1$, because the potential becomes steep as
 $\beta \to 1$ \footnote{The KKLMMT model \cite{Kachru:2003sx} corresponds to $\beta =0$.}.
The inflationary footprint is
\begin{equation}
n_s -1 \sim \frac{2 \beta}{3} \left(1 - \frac{5}{(e^{2\beta \N}-1)} \right)
\end{equation}
with $r(\beta \sim 0.1) \sim 10^{-4}$, significantly larger than the KKLMMT model where $r\sim 10^{-9}$ \cite{Kachru:2003sx}. 
The full phenomenology is discussed in \cite{Firouzjahi:2005dh} 
where they used the WMAP data to bound the parameter $\beta$, however for fully UV complete scenarios we 
expect this to be fixed.

Finally we mention a $D$-term inflationary model in IIA, arising from the intersection of four brane stacks 
in a phenomenological configuration \cite{Dutta:2007cr}. 
The inflaton connects two different brane stacks, and has a one-loop potential of the form
\begin{equation}
V(\phi) = g^2 \xi^2 \left(1 + \frac{g^2}{4 \pi^2} \ln \left( \frac{\lambda \phi^2}{\Lambda^2}\right) \right)
\end{equation}
with scalar index $n_s = 1 - 1/\N^*$ 
(this is the $p=0$ small field example of Section \ref{subsec:dis}). The FI-term $\xi$ sets 
the scale for the power spectrum, and the WMAP normalisation imposes $\xi \sim (10^{15} \rm{Gev})^2$ 
assuming $g^2 \sim 10^{-2}$. Any cosmic strings formed in this process have a tension 
$G \mu \sim \xi M_p^{-2}$, which is fortunately well below the current observable threshold.

{\bf \it Axion monodromy.} An interesting proposal which has developed 
from the brane models we have been discussing relies on axion monodromy \cite{Silverstein:2008sg}. 
This requires a $D5$-brane to be present in a type  IIB compactification, wrapping
some two-cycle ($\Sigma_2$) and carrying NS-NS flux on the worldvolume \footnote{An S-dual system involving NS$5$-branes can also be constructed, however the axion
now arises from integration of the RR flux.}. 
One can associate an axion to this flux through a term $b = 2\pi \int_{\Sigma} B$. Computation
of the brane action in a particular compactification, results in a scalar (inflaton) potential 
that is linear in $b$ (provided that it is larger than the size of the compact cycle), and therefore
gives rise to a linear inflaton potential of the form of Eq.(\ref{examples}) (b) with $p=1$. 
Such a potential is strongly disfavored 
from a field theory perspective, and therefore could be considered a 
signature of stringy physics. Since a relation between $r$ and $n_s$ 
exists for this model it is testable without knowing $\N^*$. For $\N^* \sim 60$, one finds $r \sim 0.07, n_S \sim 0.975$ \cite{Silverstein:2008sg}, 
within current WMAP bounds.
Compactification of the model using $D4$-branes on a Nil-manifold results in a fractional 
power law potential of the form $\phi^{2/3}$, the predictions of which follow from  Eq.(\ref{examples}) (a) with $p=2/3$.  
Again, such models are disfavored in field theory\footnote{One caveat here is that linear (and fractional) models can be found in the 
SUGRA literature \cite{Takahashi:2010ky}
and therefore the sub-leading corrections present in the axion monodromy framework will be important in breaking the degeneracy with such models.}. 
One other interesting feature is that, dependent on the 
details of the compactification, the potential may have a superimposed small oscillation from instanton effects, which 
would lead to an oscillatory feature in the power spectrum \cite{Flauger:2009ab} and more pronounced oscillatory 
features in the bi-spectrum \cite{Chen:2008wn,Hannestad:2009yx}. If, for example, the level of the effect was too small to affect the 
power-spectrum relations discussed above, but could be seen in the bi-spectrum, this combined evidence would be 
powerfully predictive.


{\bf \it Tachyon models.} One of the simplest, and most popular, models is that of tachyon inflation, 
driven by the condensation of an open string mode on a non-BPS $D$-brane \cite{Sen:2003mv}. 
Early constructions were unable to satisfy observational bounds because the tachyon mass was too large, however
once warped models were developed this constraint could be evaded \cite{Kofman:2002rh, Choudhury:2002xu}. Although the action is non-linear, 
tachyon inflation does not generate large $f_{nl}$ because inflation ends before (ultra)-relativistic effects become dominant. 
A step towards a concrete UV embedding of this theory was developed in \cite{Cremades:2005ir, Panda:2007ie} where they considered a non-BPS $D6$ in a geometry
generated by $D3$ flux. The scalar index was found to be
$0.94<n_s < 0.97$ for a string coupling in the range $0.1<g_s <0.34$, which suggested that larger $D3$-flux would lead to better agreement with
experiment. 

{\bf \it Non-local Inflation} Thus far the models we have considered have been in the context of low energy supergravity.  
In the case of $D$-brane and tachyon actions we have considered models which contain terms of higher order in $X$, but 
none of the models retains higher derivatives. Generally there will be an infinite tower of such higher
derivatives which, at energies above the string scale, cannot be ignored. 
A radical proposal, referred to as non-local inflation, aims to study the effect of such a tower in a limited way.
One example uses the action for the tachyon from a toy model of string theory, the P-adic string, where the world-sheet coordinates are restricted to the 
set of P-adic numbers\footnote{This assumption was argued to be relaxed in to consider any positive integer $(p)$ \cite{Freund:1987kt}.}. Other 
settings include the action for the tachyon derived from truncated cubic string field theory (CSFT) (see for example \cite{Arefeva:2001ps}), 
which can also only be considered as a 
toy model. If non-local effects are generic, studying these models {\it may} 
still tell us something interesting about possible stringy observables.  
A general non-local scalar field action takes the form ${\cal L}_{\phi} = \phi G(\Box)\phi-V(\phi)$, where $\Box$ is the d'Alembertian operator, and $G$ 
an arbitrary analytic function. Considering this Lagrangian, one discovers that inflation can proceed even if $V$ is naively 
too steep for slow-roll to be supported,  in a sense the additional derivatives act as friction terms.
For the p-adic string, $G(x)=-\gamma^4 \exp(-\alpha x)$, and $V(\phi) = \gamma^4 \phi^{p+1}/(1+p)$ where 
$\alpha = \ln p/(2m_s^2)$ and $\gamma^4 = (m_s^4 /g_s^2 )(p^2 /(p-1))$. 
The potential is naively too steep for large $p$, however the effects of the infinite number of derivatives leads to a 
consistent dual hilltop inflationary model \cite{Barnaby:2006hi, Lidsey:2007wa}. 
While it is too early to say that inflation is more natural 
once higher derivative effects are taken into account, it is an intriguing possibility. 
A final remark is that initial calculations suggest this model can give rise to large non-Gaussianity, with 
\cite{Barnaby:2006hi}
\begin{equation}
f^{\rm eq}_{nl} = \frac{5(\N-2)}{24 \ln p} \sqrt{p r}|n_s-1|^{3/2} \frac{(p-1)}{(p+1)}, \hspace{0.4cm} r = \frac{(p+1)}{2p}|n_s-1|e^{-\N |n_s-1|}
\end{equation}
where $r$ decreases for larger values of $p$, $r$ is unobservably small of order $\mathcal{O}(10^{-3})$, $f_{nl}$ clearly scales as $\sqrt{p r}$, 
and for large $p,r$ becomes independent of $p$, allowing $\fnl$ to become large. 
One issue is that determining the precise end of inflation (and hence precise observational parameters) requires knowledge of the 
dynamics of the fully non-linear regime, which is extremely difficult \cite{Mulryne:2008iq}.
Interestingly the shape of the non-Gaussianity is different from DBI models, peaking
on squeezed triangles, similar to the shape produced by multi-field models.  
This signature is interesting, but whether 
it can be distinguished from the multi-field models is unclear, and it may require knowledge of higher order statistics such as the trispectrum.
The scalar index is red with current calculations giving
\begin{equation}
|n_s-1| \sim \frac{4}{3} \left( \frac{m_s}{H}\right)^2
\end{equation}
and $H> m_s$. Such a condition is acceptable in this model because of the possible UV completion and, as discussed, 
is the source of the novel features present.

{\bf \it Assisted inflation models.}  A number of the models above may contain multiple fields 
when we move beyond the minimal scenarios.
Typically a small number of fields are considered, both in order to keep the calculation tractable 
and because a larger number 
of fields implies more freedom and hence less opportunity for models to be 
probed by observations, or make robust predictions. In the 
limit where a very large number of fields are present however, interesting effects can occur. 
In certain models; many fields with potentials 
which are naively too steep to give rise to inflation can act in a collective, assisted 
manner enabling inflation to proceed \cite{Liddle:1998jc,Kanti:1999vt}. 
This also allows each individual field to travel sub-Planckian distances. 
Furthermore the collective behaviour can appear very similar to inflation sourced by one field moving over a much 
larger distance, and hence $r$ can be large enough to be observable. From one point of 
view such scenarios appear natural, since the assisted behaviour relaxes the conditions each individual potential must satisfy, and 
 they also have the potential to be testable and predictive, because the very large number of fields can enable 
a statistical approach to making predictions.

One such model of interest is N-flation \cite{Dimopoulos:2005ac}, which considers a large number of axion fields, each paired with a modulus of the compactification, 
to act collectively to source inflation. 
With coupling neglected, each axion has a sinusoidal self-interaction potential of the form $V_n(\phi_n) =\Lambda_n^4 \left(1- \cos(2 \pi \phi_n / f_n)\right)$,
which appears like a quadratic potential 
when expanded around a minimum, with $m_n=2\pi\Lambda_n^2/f_n$. Then, if the masses are identical, the theory is effectively
that of a single field sourced by a quadratic potential of the form of Eq.(\ref{examples}) (b) with $p=2$ and hence $n_s-1 = -2/\N$, $r= 8/\N$. 
For $\N\sim 60$, and with $f<\mpl$, the number of axions required is typically thousands.  
The observational signature changes if the axion masses are not all identical, and  
a more realistic approach is to have masses distributed according to a Marcenko-Pastur probability distribution 
$p(m^2) = \sqrt{(b-m^2)(m^2-a)}/(2\pi m^2 \beta \sigma^2)$, where 
$a<m^2<b$, $a=\sigma^2 (1-\sqrt{\beta})^2$,$b= \sigma^2(1+\sqrt{\beta})^2$,  $\sigma^2 = \langle m^2 \rangle$ 
and $\beta\sim 1/2$ is typical, and depends on the dimension of the K\"ahler and complex 
moduli spaces \cite{Easther:2005zr, Piao:2006nm}.  
Remarkably this statistical approach is surprisingly testable.
When comparing to observations we must also fix initial conditions for the various fields. One approach is to also do this randomly, and one 
finds average values for the spectral index are typically lower than with equal masses, 
$n_s \approx 0.95$ for $50$ e-folds, this being insensitive 
to the distribution from which the initial conditions are drawn. $r$ is independent of the 
model parameters and given by the same expression as above
and the non-Gaussianity negligible \cite{Kim:2007bc}.  
An intriguing recent development has been the observation that when the full axion potential 
is considered, a large local $\fnl$ can be produced, with $\fnl^{\rm loc} \sim 10$ when all axions 
are taken to be identical, 
and $f= \mpl$ \cite{Kim:2010ud}, though in 
this case $r$ is negligible and $n_s$ slightly lower again, putting the model in tension 
with WMAP. This result 
has been calculated using the $\delta N$ formalism, and may be altered in the light of 
numerical simulations \cite{mulryne}, and should be 
tested in more realistic settings and mass distributions. We note that
statistical approaches may well have a role to play when confronting other complicated 
string theory models with observation.

{\bf \it M-theory models} A robust scenario in Heterotic M-theory reproduces the results of assisted power law inflation 
where the potential is exponential and  $a(t) \sim a_0 t^p$. 
Inflation here occurs before moduli stabilisation and is driven by 
the non-perturbative dynamics of $N$ five-branes along the orbifold direction \cite{Becker:2005sg}. 
Under a set of reasonable assumptions, the instanton generated scalar potential
is always the steepest direction in field space, which would be unable to support single field inflation.
The scalar index takes the expected form $n_s = 1-2/p$ where $p =N^3 + \ldots$
which is used to fix the number of branes using the WMAP data. 
Whilst the $R^4$ corrections are known, their implementation is difficult since they compete with
higher order instanton effects - spoiling the simplicity of the model. However their inclusion could break the field theory degeneracy and point to a unique
signature of M-theory. Moreover moduli stabilization and subsequent reheating in this model will no doubt further constrain the parameter space, and
test the viability of such a scenario.

A related model arises with only a single five-brane wrapped on the orbifold \cite{Buchbinder:2004nt},
where the inflaton is identified with the real part of the five-brane modulus $(x)$.
For $x \ll1$ the five-brane is localised near the visible sector, and inclusion of a FI-term in the \emph{hidden} sector uplifts
the stabilised vacuum to $dS$. Slow roll inflation (with no back-reaction) occurs in this regime with
$\N \sim \eta^{-1} \ln (x_i/x_f)$, where $x_i, x_f$ are the initial and final positions of the brane, 
and $\eta$ is the slow roll parameter which can be expressed as a ratio of the fluxes
arising from the superpotential. With $\eta =0.1, x_i = 10^4 x_f$ we find $\N \sim 80$ and $\mathcal{P}_s^2 \sim 10^{-10}$
which agrees with WMAP. Inflation ends once the five-brane dissolves into the visible sector via instanton transition, this in
turn excites vector bundle moduli resulting in a shift of the cosmological constant. 
For larger values of $x$ other moduli will be destabilised
from their vacua, and may subsequently lead to a novel inflationary footprint.

\subsection{Alternative models}

Inflation is by far the most developed, and promising, theory for the origin of structure in the 
universe. But in the context of string cosmology, other scenarios exist which could be more natural. 
We briefly mention below some attempts to develop such alternatives.

{\bf \it Ekpyrotic model.} The Ekpyrotic model is an alternative to inflation \cite{Khoury:2001wf,Lehners:2008vx,Lehners:2011kr}.  
Instead of generating perturbations during an exponential expansion,  
successive $k$ modes exit the horizon during a slow contraction. 
Predictions are made predominantly within an effective field theory, but the model can 
be embedded in Heterotic M-theory. 
In the original single field models the inflaton is associated with the distance 
between the two $5D$ `end of the world' branes located at the orbifold fixed points \footnote{The additional $6$ dimensions being compactified on a small scale.}, 
and has a steep negative potential, not directly derived from the theory. 
As the field evolves, the universe collapses and these branes approach one another. 
During the collapse the spectrum of $\zeta$ perturbations is extremely blue and not phenomenologically viable \cite{Lyth:2001pf}, however an almost
scale invariant spectrum can be produced in the Newtonian potential \cite{Gratton:2003pe}.  
Standard hot big bang cosmology is recovered when these branes collide, and it is possible that scale invariant perturbations get imprinted on $\zeta$, 
but this requires going beyond the $4$-dimensional description \cite{McFadden:2005mq} and is a potential 
weakness of the model.
An alternative suggestion is to consider a two-field model, the second field arising from the volume modulus of the internal dimensions 
\cite{Lehners:2007ac,Koyama:2007ag,Buchbinder:2007ad}. 
If both fields have steep negative potentials parametrised by $V_i= e^{-c_i(\phi) \phi}$, 
then (in field space) the shape of the potential looks like a ridge. If the 
trajectory is finely tuned such that the inflaton rolls down this ridge, then the isocurvature perturbation 
produced during collapse is close to scale invariant.  
If the trajectory subsequently curves, either by the trajectory naturally falling off the ridge or 
by `bouncing' off a boundary in 
field space, then this isocurvature perturbation can be converted into $\zeta$. 
The model predicts, 
$n_s -1 = \epsilon^{-1}(2- \partial \ln \epsilon / \partial \N)$
where the first term is blue-tilted and the second term red-tilted. 
Tensors are unobservable, which means that detection of almost
scale invariant gravitational waves will rule out Ekpyrosis, and strongly favour the simplest inflationary models.

The conversion of isocurvature to curvature perturbations has a secondary effect, which is to produce a 
large value of $\fnl$ in the squeezed shape typical of multi-field models. For the simplest case where the conversion 
occurs by naturally falling of the ridge, $f_{nl} \sim -5 c_1^2/12$, where
$1$ labels the field which dominates at late times, and can be calculated using 
the $\delta N$ formalism \cite{Koyama:2007if}. Clearly large non-Gaussianities
can be generated if $c_1 \gg 1$, and moreover $c_1\gg10$ is required for consistency of the 
spectral index with WMAP data, and therefore 
the level of non-Gaussianity is in severe tension with observation.  
In the case where the trajectories `bounce', positive and negative values of $\fnl$ are possible
 and the amplitude depends on how suddenly the bounce occurs \cite{Buchbinder:2007at,Lehners:2007wc,Lehners:2010fy}.  
Interestingly in search of a robust predictive signal, the authors have considered 
higher order statistics \cite{Lehners:2009ja}, and though no meaningful constraints currently exist, future 
observations may probe the scenario in this way. 
A final comment on this scenario is that the initial 
conditions are extremely finely tuned.  While mechanisms have been suggested to alleviate this tuning in a pre-ekpyrotic 
phase \cite{Buchbinder:2007tw}, it is hard to not to consider the evolution rather unnatural. On the other hand, because of the 
special initial conditions required to make the model work, in contrast to multi-field inflationary models, it is extremely 
testable and potentially predictive.

{\bf \it String gas cosmology.} A novel program which both aims to understand the effect 
of the extended nature of strings on the early universe, and has attempted to replace inflation with an alternative mechanism of 
generating scale invariant perturbations, is that of string gas cosmology \cite{Battefeld:2005av, Brandenberger:2008nx}. 
This involves coupling the graviton and dilaton to a string gas 
(which may also include other degrees of freedom such as branes) and using T-duality to interchange winding and momentum modes.
The model predicts a slightly red scalar index, but a blue tilt for gravitational waves which allows the theory to be ruled out. 
Interestingly the theory favours the Heterotic string due to existence of enhanced symmetries necessary for moduli stabilization.

{\bf \it Pre-big bang.} Older models of the early (stringy) universe restricted themselves purely to the dilaton sector, at leading order in 
world-sheet and string loops (with $V(\phi)=0$). 
Application of generalised T-duality led to the existence of scale-factor duality
which aimed to resolve the Big Bang singularity by replacing it by an epoch of high, but finite, curvature \cite{Gasperini:2007ar}. 
At early times, before this 'Big Bang', we find $g_s \ll 1$
allowing us to probe the perturbative string vacuum without worrying about loop corrections. The coupling increases as we pass through this
singularity until it becomes constant at late times. 
However this simple picture does not account for the observed perturbations, the dilaton perturbations leading to a strongly blue spectrum, instead one must
consider a curvaton mechanism driven by an axionic field dual to $B_{\mu nu}$. In turn this drives the production of both a graviton and dilaton background, where
the dilaton mass can be $m \ge 10^{-23}eV$, which is detectable (in principle). The curvaton potential is assumed to be quadratic, in which case the predictions
are the same as canonical $m^2 \phi^2$ inflation and satisfy the WMAP data.
Interestingly the type I string is favoured over the Heterotic string in such models due to the difference in primordial magnetic seed production.

\subsection{Reheating}

Reheating is a significantly less developed topic in comparison to inflationary model building, but hugely important. 
Indeed in many instances there are only vague ideas as to the existence/location of the 
standard model sector. As we have already discussed this lack of post-inflationary knowledge 
means $\N$ is not fully determined, and hence observational predictions ambiguous. Reheating 
is also interesting in its own right, and 
although the energy scale involved 
is significantly smaller than that associated with inflation, one may still hope that there is
sensitivity to the extended nature of the superstring.

A landmark paper \cite{Frey:2005jk} considered the case of (warped) closed and open string reheating. 
The closed string sector was purely in the supergravity limit, and despite leading to interesting results, they argued 
that it was hard to distinguish between string theory and Kaluza-Klein physics unless
one could examine the $H/M$ expansion order by order. In the open string case, the strings were 
argued to redshift like matter. Both approaches further suggested the formation of long string
networks during inflation, which could be detectable in the CMB.

More recent papers discuss the case of K\"ahler inflation in two different classes of closed
string model \cite{Cicoli:2010ha}, depending on whether the inflaton is the size of a blow-up mode of the Calabi-Yau (BI model), 
or whether it is the size of the $K3$ fibre in Fibre inflation (FI model). Both cases involve the leading order
$\alpha'$ corrections, although $g_s$ corrections are argued to be decoupled from the theory. 
The results indicate that there will always be hidden sectors present, that the BI model 
requires a much higher level of fine-tuning than FI since the hidden sector must wrap the
same four-cycle as the inflaton in the former scenario - however despite the higher degree of tuning, the 
FI scenario leads to a small reheat temperature which is incompatible with TeV scale SUSY. As such, it should be disfavored. 
The final conclusion was that the hidden 
and visible sectors are not directly coupled, and the degrees of freedom in the visible sector
cannot be more strongly coupled than its hidden sector counterparts. 
They further identified two generic problems with such a reheating mechanism; i) Inflationary energy will be transferred to the 
hidden sector. This is not a problem if the hidden sector degrees of freedom are relativistic, but does
require a curvaton mechanism to generate the perturbations in the visible sector. ii) There
may be overproduction of hidden sector dark matter, which would spoil BBN.

Reheating of the axion monodromy scenario has also been explored, at least in the IIA framework where the $D4$-brane unwinds \cite{Brandenberger:2008kn}. The
$D4$-brane passes through a $D6$-brane, however the open string modes present during the collision act as a braking force. This
interaction was described by a toy field theory model, and suggested that no energy was transferred during these collisions. All
the reheating energy is dumped into the final collision event, resulting in a high reheating temperature. However backreactive effects
were not considered, and may spoil the simple field theory picture.

\subsection{Cosmic Strings}
A striking prediction of several string scenarios is the formation 
of cosmic super-strings during or even \emph{after} 
inflation \cite{Copeland:2003bj}. 
Such objects confront observation in a number of ways. First they contribute to primordial density 
perturbations, and CMB analysis 
indicates that they cannot account for more than $11\%$ of the power \cite{Bevis:2007qz}, 
limiting their tension to $G\mu<2.1\times10^{-7}$ \cite{Fraisse:2006xc}. Moreover, 
the vector-mode perturbations they source will lead to CMB polarisation 
potentially observable by the Planck satellite or future missions \cite{Pogosian:2007gi}. 
They can strongly gravitationally lens background objects in a distinctive 
way, but as yet no candidate lensing event has been seen, and through vector perturbations they will 
rotate images observed in weak lensing surveys \cite{Thomas:2009bm}. The most promising way in which they can be detected, however,  
is through a gravitational wave signal produced by cusps on the strings, potentially 
detectable for tensions as low as $G\mu \sim 10^{-10}$, though whether this signal will be 
observable by LIGO, LISA or BBO is model dependent \cite{Polchinski:2007qc}. Current limits come from pulsar timing bounds, 
which lead to $G\mu<1.5\times 10^{-8} c^{-3/2}$, where $c$ is the number of cusps per string loop.
 
The discovery of evidence for cosmic super-strings would be extremely powerful evidence for string theory. 
This requires, however, that they be distinguished from standard cosmic strings, and this is 
extremely challenging (see for example \cite{Copeland:2009ga}). One important difference 
is that intersecting field theory strings recombine with probability $P$, passing through one another with 
probability $1-P$. Numerical simulation, along with theoretical calculation, suggests
that $P \sim 1$ to a remarkably high degree for strings of the same type. For F-F strings it turns out 
that $P \sim g_s^2$, suggesting they will pass through one another rather than reconnecting. 
If  $P$ and $\mu$ can be determined through gravitational wave observation, these possibilities can be distinguished. 
For F-D or D-D interactions, the value of $P$ is 
less constrained, valued in the range $P \in 10^{-3} \ldots 1$. Therefore the perturbative 
F-F interaction could provide the best direct evidence for string theory \cite{Copeland:2003bj, Sakellariadou:2008ie}. 
Cosmic super-strings of different type may also combine to form $(p,q)$-strings or 
even networks \cite{Schwarz:1995dk, Copeland:2006if}. 
Such objects have a distinctive tension spectrum which is remarkably
difficult to recreate using perturbative field theory.

In Heterotic M-theory one can consider three different types of cosmic strings \cite{Gwyn:2008fe}, 
a membrane wrapped on the $x^{11}$ direction, a five-brane wrapped on a four-cycle $(\Sigma_4)$
of the internal space or a five-brane wrapped on the product space $\Sigma_3 \times x^{11}$ - where $\Sigma_3$ is a three-cycle. Such 
strings are formed after the inflationary phases discussed in \cite{Becker:2005sg, Buchbinder:2004nt}. 
The membrane tension is too large and is strongly dis-favoured. The only stable five-brane string is the one wrapped on $\Sigma_4$ because
one must turn on a gauge field to cancel the anomaly term, which must live on the boundary and only the brane wrapped on $\Sigma_4$ can be stabilised. Such
a string can be superconducting and generates seed magnetic fields that are coherent on \emph{all} cosmological scales. 
Those fields at large scales cannot be amplified by a dynamo mechanism and therefore will have a weak, coherent amplitude, 
which may be detectable in future experiments.

\section{Discussion}\label{sec:Discuss}
In this paper we have provided a critical, non-exhaustive, review of the current observational status of string theory using
cosmological data. We have emphasised that inflationary model building has been a success, in the sense that  
string theory can accommodate inflation, and 
that the footprints of many
different models are testable and conform with WMAP data. Planck will offer a 
considerably more stringent test, and will undoubtedly rule out many models. 
Most models, however, exist as 
field theories and do not make 
direct predictions for how stringy corrections lead to shifts in observables. Never-the-less we have 
reviewed a number of signals, such as non-Gaussianity, and relations between parameters, which might in combination 
with other considerations be evidence 
for stringy models. There are even some indications that inflation might be more natural 
in certain stringy settings. Alternatives to inflation are less well developed, but may also offer predictions which allow them to be probed observationally. 

Reheating in inflationary models which includes stringy corrections
will be different to those in field theory, and have been examined in several specific instances with important phenomenological consequences. However
much more work needs to be done, particularly in M-theory, if these models are to be falsifiable.
The cleanest signal for string theory still remains the detection of a cosmic superstring through the colliding F-F channel, 
although more work needs to be done on understanding
how such strings scale during the reheating phase, and whether the dynamics of network formation is likely to be important.

We hope our review has highlighted the need for future work in a number of key areas. 
First, we note that parts of a model are often 
studied and compared with observation 
in isolation. For example inflationary predictions, and subsequent evolution including the reheating scale are generally treated  
separately, while in reality the later impacts on the former through $\N^*$. Likewise the production of cosmic strings is 
often treated separately and compared with observation independently of inflationary constraints, while 
the presence of both will alter the constraints considerably. This points to the need for a more holistic treatment, and for 
work on complicated questions. Another key issue is correctly accounting for the 
presence of more than one inflationary field, typical of complex models. Perhaps most importantly is the 
issue of the inclusion of higher order corrections, leading to robust 
cosmological tests.

Aside from these issues there exist other ways in which string theory could be cosmologically tested. 
One proposal is to embed inflation in non-geometric flux compactifications. Since non-geometric fluxes arise from multiple T-dualities, they
are inherently stringy, therefore observables should directly probe the underlying theory. Work along these lines \cite{Flauger:2008ad}
determined a no-go theorem for massive IIA with metric flux. Only the $\mathbb{Z}_2 \times \mathbb{Z}_2$ case evaded this
stringent theorem, but did not lead to slow-roll inflation. Future models are likely to evade this theorem, and it will be interesting to determine their
footprints.

Finally, we note that there are many other areas of string cosmology we have not been able to discuss, such as corrections to the power spectra due to new physics 
at high energy scales \cite{Danielsson:2002kx}, or effects from 
large extra dimensions. It is likely that work on several fronts will be necessary to determine which
footprints are inherently stringy. Although thus far there are no truly convincing models or models with signals uniquely 
predictive of string theory, there remains much work to be done and the future for this field remains bright.

\section*{Acknowledgments}
DJM is supported by the Science and Technology Facilities Council grant ST/H002855/1. We would like to 
thank Reza Tavakol and Xingang Chen for useful comments.

\end{document}